\documentclass[a4paper,12pt]{article}
\usepackage{epsfig}
\date{\today}
\begin{document}

\renewcommand{\thefootnote}{\fnsymbol{footnote}}

\rightline{}
\vskip 1cm
\begin{center}
{\bf \large{  CP--Violating Invariants in Supersymmetry \\[10mm]}}
{ Oleg Lebedev \\[6mm]}
\small{Centre for Theoretical Physics, University of Sussex, Brighton BN1
9QJ,~~UK\\[2mm]}
\end{center}

\hrule
\vskip 0.3cm
\begin{minipage}[h]{14.0cm}
\begin{center}
\small{\bf Abstract}\\[3mm]
\end{center}

I study the weak basis CP-violating invariants in supersymmetric models, in particular those which
cannot be expressed in terms of the Jarlskog--type invariants, and find basis--independent conditions
for CP conservation. With an example of the $K - \bar K$ mixing, I clarify what are the combinations of
supersymmetric parameters which are constrained by experiment.

\end{minipage}
\vskip 0.7cm
\hrule
\vskip 1cm

\section{Introduction.}

The Standard Model possesses only one CP--odd quantity invariant under a quark basis transformation
(apart from $\bar \theta_{\rm QCD}$), which is known as the Jarlskog invariant \cite{Jarlskog:1985ht}:
\begin{eqnarray}
J&=&{\rm Im~\biggl(~ Det} \left[ Y^u Y^{u\dagger}, Y^d Y^{d\dagger}  \right]~
\biggr) \nonumber\\
&\propto& (m_t^2-m_u^2)(m_t^2-m_c^2)(m_c^2-m_u^2)
          (m_b^2-m_d^2)(m_b^2-m_s^2)(m_s^2-m_d^2)\nonumber\\
&\times&         {\rm Im}(V_{11}V_{22}V_{12}^* V_{21}^*)\;,
\label{jar}
\end{eqnarray}
where $Y_{ij}^a$ are the Yukawa matrices and $V_{ij}$ is the CKM matrix. 
In supersymmetric models, there are many additional sources of CP violation as well as new flavor structures \cite{Haber:1984rc}. In this paper, I will concentrate on the quark--squark sector and will ignore leptonic effects for simplicity. Then,
the relevant superpotential and the soft SUSY breaking terms are written as follows:
\begin{eqnarray}
\Delta W&=&-\hat{H}_2 Y^u_{ij} \hat{Q}_i \hat{U}_j 
+\hat{H}_1 Y^d_{ij} \hat{Q}_i \hat{D}_j 
-\mu\hat{H}_1\hat{H}_2 \;, \nonumber\\
\Delta V_{\rm s.b.} &=& M^{2 q_{_L}}_{ij} \tilde q_{Li} \tilde q_{Lj}^*+
                 M^{2 u_{_R}}_{ij} \tilde u_{Ri} \tilde u_{Rj}^*+
                 M^{2 d_{_R}}_{ij} \tilde d_{Ri} \tilde d_{Rj}^* + m_1^2 \vert H_1 \vert^2 
                                                                 + m_2^2 \vert H_2 \vert^2 
 \nonumber \\ & + &
                 \left( -H_2 A^u_{ij} \tilde q_i \tilde{u}_j^* +
                 H_1 A^d_{ij} \tilde q_i \tilde{d}_j^* 
                   - B\mu H_1 H_2 + {\rm h.c.} \right)
                 -{1\over 2} \sum_i M_i \lambda_i \lambda_i . 
\label{l}
\end{eqnarray}
Clearly, all quantities with flavor indices can contain flavor--dependent and, possibly, flavor--independent
CP violating phases. The latter (e.g. overall phases of the A-terms, $\mu$, $B\mu$, $M_i$) 
have been studied scrupulously in the past, while 
the former  (e.g. the off--diagonal phases of $M^{2 q_{_L}}$, etc.)
have not received as much attention. One of the reasons besides cumbersomeness
is that such phases are basis--dependent and thus should be treated with care.
In the case of the Standard Model, the flavor--dependent CKM phase can be expressed in terms 
of the Jarlskog invariant. An important question to address is what is the generalization of the
Jarlskog invariant for supersymmetric models and how it is related to the SUSY CP phases.

The class of supersymmetric models under consideration possesses the following symmetries
\begin{equation}
U(3)_{\hat Q_L} \times U(3)_{\hat U_R} \times U(3)_{\hat D_R}
\label{symmetry}
\end{equation}
acting on the quark superfields, which preserve the structure of the supergauge interactions.
The CP transformation acts on the Yukawa and mass matrices as the complex conjugation:
\begin{equation}
M \stackrel{CP}{\longrightarrow} M^*\;,
\end{equation}
where $M=\{ Y^u,Y^d,M^{2 q_{_L}},M^{2 u_{_R}},M^{2 d_{_R}},A^u,A^d \}$. 
If this can be ``undone'' with the  symmetry transformation (\ref{symmetry}), the $physical$ flavor--dependent CP phases vanish. Supersymmetric models also possess the Peccei-Quinn and R symmetries $U(1)_{\rm PQ}$ and 
$U(1)_{\rm R}$, which allow us to eliminate two of the flavor--independent phases (see e.g. \cite{Abel:2001vy}). 
Then, in order for CP to be conserved, the invariant CP--phases
\begin{equation}
{\rm Arg}\left[  (B\mu)^* \mu M_i  \right] \;\;,\;\; {\rm Arg}\left[A^*_\alpha M_i  \right]
\end{equation}
have to vanish too. Here Arg$\left[ A_\alpha \right], \alpha=u,d$, denotes the ``overall'' phases of 
the A-terms which can be defined in a basis--independent way as
\begin{equation}
{\rm Arg}\left[ A_\alpha \right] \equiv {1\over 3} {\rm Arg} \left( {\rm Det }\left[ A_\alpha Y^\dagger_\alpha   \right] \right) \;,
\end{equation}
provided this determinant is non--zero.
The flavor--independent phases are not affected by the quark superfield  basis transformation and 
thus are physically meaningful. The discussion of the flavor--dependent phases is much more involved.
The main subject of this paper is to find the $physical$ 
CP--phases, i.e. those which are invariant under phase redefinitions of the quark superfields
in analogy with the CKM phase,
and  the corresponding basis--invariant quantities similar to the Jarlskog invariant. 

The paper is organized as follows. In section 2 I build up necessary techniques
to handle the issues of CP violation in theories with many flavor structures.
In section 3 I apply these methods to the Minimal Supersymmetric Standard Model and 
provide some examples of how observable quantities can be written in manifestly 
reparametrization invariant form.

\section{Auxiliary Construction. }

\subsection{The case of three matrices.}

In the Standard Model, the CP--odd invariant is built on the hermitian quantities $Y^u Y^{u\dagger}$ and $Y^d Y^{d\dagger}$
which transform in the same way under a basis transformation, i.e.  
$Y^u Y^{u\dagger} \rightarrow  U_L Y^u Y^{u\dagger} U_L^\dagger$ and 
$Y^d Y^{d\dagger} \rightarrow  U_L Y^d Y^{d\dagger} U_L^\dagger$.
Suppose, in addition to these, we have another quantity with the same transformation property, for instance, 
$M^{2 q_{_L}}$.
Denoting $A\equiv Y^u Y^{u\dagger}$, $B \equiv Y^d Y^{d\dagger}$, and $C \equiv M^{2 q_{_L}}$, we have 
\begin{eqnarray}
&& A \rightarrow U_L ~A~ U_L^\dagger \;\;,\;\; B \rightarrow U_L ~B~ U_L^\dagger \;\;,\;\; C \rightarrow U_L ~C~ U_L^\dagger\;,
\end{eqnarray}
where $U_L$ is a $U(3)$ quark superfield transformation $\hat Q_L \rightarrow (U_L)^T \hat Q_L$ (clearly, these quantities are invariant under the right--handed superfield transformations). 
What are the invariant CP--violating
quantities and the physical CP--phases in this case?

Taking advantage of  the unitary symmetry, let us go over to the basis where one of the matrices, 
say $A$, is diagonal. In this basis,
\begin{eqnarray}
{A= \left(      \matrix{a_1 & 0 & 0 \cr
                       0 & a_2 & 0 \cr
                       0 & 0  & a_3  }   \right) \;,\;
B=\left(      \matrix{b_{11} &  b_{12} &  b_{13} \cr
                        b_{12}^* & b_{22} &  b_{23} \cr
                        b_{13}^* & b_{23}^*  & b_{33}  }   \right) \;,\;
C=\left(      \matrix{c_{11} & c_{12} & c_{13}   \cr
                      c_{12}^* & c_{22} & c_{23} \cr
                      c_{13}^* & c_{23}^*  & c_{33}  }   \right) \;.}  \label{ABC}
\end{eqnarray}
The residual symmetry is associated with the $U(3)$ generators commuting with the diagonal matrix $A$. These are 
two  $SU(3)$  Cartan subalgebra generators and the generator proportional to the unit matrix. This means
that $B$ and $C$ are defined up to the phase transformation $B,C \rightarrow U_1~ B,C~  U_1^\dagger$ with
\begin{equation}
U_1= {\rm diag}\left( e^{i \delta_1}, e^{i \delta_2}, e^{i \delta_3} \right) \;.
\end{equation}
Under this phase transformation the matrix elements  transform as
\begin{equation}
b_{ij} \longrightarrow b_{ij}~e^{i(\delta_i-\delta_j)}
\end{equation}
and similarly for $c_{ij}$. 
Physically this freedom corresponds to the arbitrariness in the choice of the quark superfield phases.
The quark and squark fields are to be transformed with the same phases in order not to pick up 
CP phases in the interaction vertices such as $\tilde q^* q  \tilde g$.
In our basis,
$A={\rm diag}(m_u^2,m_c^2,m_t^2)/v_2^2$, $B=V{\rm diag}(m_d^2,m_s^2,m_b^2)V^\dagger/v_1^2$, where $V$ is the CKM matrix
and $v_{1,2}$ are the Higgs VEVs. 
The supergauge vertices are diagonal and the flavor mixing is contained in the propagators 
(and the non--gauge vertices). 
The residual rephasing symmetry implies
that all physical quantities must be invariant under a phase redefinition of the quark superfields.

If we have only two matrices $A$ and $B$, the only reparametrization--invariant CP phase   we can construct
is the CKM--type phase
\begin{equation}
\phi_0 = {\rm Arg}(b_{12} b_{13}^* b_{23})\;.
\end{equation}
In the case of three matrices, there are 3 additional {\it non--CKM--type} invariant phases
\begin{equation}
\phi_i =  \epsilon_{ijk}~ {\rm Arg}(b_{jk} c_{jk}^*)\;,
\end{equation}
i.e. Arg$(b_{12} c_{12}^*)$, etc. In the non--degenerate case, 
the other physical CP--phases can be expressed in terms of these 4 phases.
For instance, the CKM--type phase for the matrix C,  $\phi_0' \equiv$Arg$(c_{12} c_{13}^* c_{23})$,  is 
given by $\phi_0'=\phi_0-\phi_1-\phi_2-\phi_3$  (yet, this is not true in the degenerate case, e.g.
when some $b_{ij}=0$).  
The number of physical phases can also be computed by a simple 
parameter counting. Three hermitian matrices have  nine phases. A $U(3)$ transformation has six phases,
of which one leaves all hermitian matrices invariant. Thus the number of non--removable phases is 9-5=4.
An interesting feature here is that a new class of reparametrization--invariant CP--phases, not expressible
in terms of those of the CKM type, arises.

Consequently, the necessary and sufficient conditions for CP conservation are
\begin{equation}
\phi_0= \phi_0' = \phi_i =0 \;\; {\rm mod}\;\; \pi \;\;\; (i=1,2,3)\;.
\end{equation}

Having identified a set of the physical CP phases, one may ask what are the weak basis CP--odd invariants
associated with such phases. In the case of two matrices, there is only one independent CP--odd invariant
which can be written as
\begin{equation}
J_{AB}= {\rm Im}  {\rm Tr [A,B]^3}  \;.
\end{equation}
It is proportional to the Jarlskog invariant of Eq.(\ref{jar}) and $\sin\phi_0$.
In the case of three matrices with the same transformation properties, one can construct a number of CP--odd
invariants such as\footnote{The commutator under the trace can also be raised to an odd power.
This will not provide independent invariants and I omit its discussion for brevity.} 
\begin{equation}
K_{ABC}(p,q,r)={\rm Im}  {\rm Tr} [A^p,B^q]C^r       
\end{equation}
with integer $p,q,r$. This invariant can also be written as ${\rm Im}  {\rm Tr} A^p [B^q,C^r]$  
or an imaginary part of the trace of the completely
antisymmetric product of $A^p,B^q$, and $C^r$. For $p=q=r=1$ it is proportional to a linear combination of 
$\sin \phi_i$ ($i=1,2,3$):
\begin{equation}
K_{ABC}(1,1,1)=2(a_1-a_2) \vert b_{12} c_{12} \vert \sin \phi_3 +
               2(a_2-a_3) \vert b_{23} c_{23} \vert \sin \phi_1 +
               2(a_3-a_1) \vert b_{13} c_{13} \vert \sin \phi_2 \;.
\end{equation}
It is worth emphasising that the $K$-invariants are entirely $new$ objects 
which $cannot$ be expressed in terms of the Jarlskog invariants and vice versa. 
The simplest way to see that is to 
imagine that $A$ and $B$ (and maybe $C$) have 2 degenerate eigenvalues. Then all Jarlskog invariants 
$J_{AB}$, $J_{BC}$, and $J_{CA}$ vanish. Yet, the $K$-invariants can be nonzero. And conversely,
suppose that $A$ is proportional to the unit matrix. Then all $K$-invariants vanish, while 
$J_{BC}$ can be nonzero.

It is instructive to express these invariants in terms of the eigenvalues 
and the mutual CKM--type matrices. That is,  write Eq.(\ref{ABC}) as
\begin{equation}
A= {\rm diag}(a_1,a_2,a_3)\;,\; 
B= V ~{\rm diag}(b_1,b_2,b_3)~ V^\dagger \;,\;
C= U ~{\rm diag}(c_1,c_2,c_3)~ U^\dagger \;.
\end{equation}
Then,
\begin{equation}
K_{ABC}(p,q,r)= \sum_{ijkl} (a^p_i-a_j^p) b_k^q c_l^r V_{ik} V_{jk}^* U_{jl} U_{il}^* \;.
\end{equation}
Note the appearance of the rephasing invariant quantities $V_{ik} V_{jk}^* U_{jl} U_{il}^*$
which generalize the CKM--type combination $V_{ik} V_{jk}^* V_{jl} V_{il}^*$. To be exact,
there are three independent invariant quantities 
\begin{eqnarray}
&& \phi_1= {\rm Arg } \left( \sum_i b_i V_{2i} V_{3i}^* \right) \left(\sum_i c_i U_{2i} U_{3i}^* \right)^* \;, \nonumber\\
&& \phi_2= -{\rm Arg } \left(\sum_i b_i V_{1i} V_{3i}^* \right) \left(\sum_i c_i U_{1i} U_{3i}^* \right)^* \;, \nonumber\\ 
&& \phi_3= {\rm Arg } \left(\sum_i b_i V_{1i} V_{2i}^* \right) \left(\sum_i c_i U_{1i} U_{2i}^* \right)^* \;, 
\end{eqnarray}
while the other are functions of these and $\phi_0$. Indeed, let us consider an example of $K_{ABC}(1,2,1)$.
The arising invariant quantity is, e.g.
\begin{equation}
{\rm Im }\left(\sum_i b_i^2 V_{1i} V_{2i}^* \right) \left(\sum_i c_i U_{1i} U_{2i}^* \right)^*\;.
\label{example}
\end{equation}
Rewriting 
\begin{equation}
\sum_i b_i^2 V_{1i} V_{2i}^* = \sum_{ijk} \left( b_i V_{1i} V_{ki}^*   \right) \left( b_j V_{kj} V_{2j}^*   \right)\;,
\end{equation}
and extracting terms with different $k$,
Eq.(\ref{example}) can be brought to the form
\begin{eqnarray}
&&  \sin\phi_3  \sum_i b_i \left[  \vert V_{1i}\vert^2 + \vert V_{2i}\vert^2 \right] 
 \left\vert \sum_i b_i V_{1i} V_{2i}^*\right\vert     \left\vert \sum_i c_i U_{1i} U_{2i}^*\right\vert +                      \
\nonumber\\ 
&&  \sin(\phi_3 - \phi_0)   \left\vert \sum_i b_i V_{1i} V_{3i}^* \right\vert 
    \left\vert \sum_i b_i V_{3i} V_{2i}^* \right\vert  \left\vert \sum_i c_i U_{1i} U_{2i}^*\right\vert    \;.    
\end{eqnarray}
This can be seen even more easily in terms of the original matrix elements $b_{ij}\equiv \sum_k b_k V_{ik} V_{jk}^*$
and $c_{ij}\equiv \sum_k c_k U_{ik} U_{jk}^*$. 

We therefore see that, as expected, not all of the $K$-invariants
are independent. From the above exercise it is clear  that, in the non-degenerate case, one can choose  
\begin{equation}
J_{AB}\;,\; K_{ABC}(1,1,1)\;,\; K_{ABC}(2,1,1) \;,\; K_{ABC}(1,2,1)
\label{basis}
\end{equation}
as the basis of the CP--odd invariants. Given the values of these four invariants, the mixing angles, and
the eigenvalues, one can solve for the physical CP--phases $\phi_i$ ($i=0..3$). 
A complication here, compared to the Standard Model, is that the $K$--invariants  generally  are non--trivial functions
of these CP--phases.
The necessary and sufficient conditions for CP conservation can be expressed as
\begin{equation}
{\rm Im}  {\rm Tr [A,B]^3}={\rm Im}  {\rm Tr [B,C]^3}={\rm Im}  {\rm Tr [C,A]^3}=
{\rm Im}  {\rm Tr [A^p,B^q]C^r} =0
\end{equation}
for any $p,q,r$.  

\subsubsection{Degenerate case.}

So far we have been assuming that all of the physical phases are non--zero. It is important to find out under which
circumstances some of them vanish. To do that, let us go back to Eq.(\ref{ABC}). Now, suppose that two eigenvalues
of matrix $A$ are degenerate, $a_1=a_2$. In this case, one of the physical phases will disappear.
Indeed, the residual symmetry in this case is $U(2)\times U(1)$. Using this symmetry the upper left 2$\times$2 block
of matrix B can be diagonalized:
\begin{equation}
   U_2 \; \left( \matrix{  b_{11} & b_{12}  \cr
          b_{12}^* & b_{22} }  \right)  \; U_2^\dagger  \;\;\;\; \longrightarrow \;\;\;\; 
\left( \matrix{  b_{11}' & 0 \cr
          0 & b_{22}' } \right)\;,
\end{equation}
where $U_2$ is a $U(2)$ matrix.
As a result, the invariant phase $\phi_3$ vanishes. Of course, the CKM--type phase $\phi_0$ also vanishes,
but it can be replaced with an analogous phase built from the matrix elements of $C$:
$\phi_0' = {\rm Arg}(c_{12} c_{13}^* c_{23})$.  Thus, the basis for the remaining physical phases can be chosen as
\begin{equation} 
\phi_0' \;,\; \phi_1 \;,\; \phi_2 \;.
\end{equation}

Now suppose that two of the eigenvalues of matrix $B$ are also degenerate. By a unitary (permutational)
transformation these can be made $b_1$ and $b_2$. 
This introduces an additional $U(2)$ symmetry, so
the CKM matrix between $A$ and $B$ is now 
defined only up to a $U(2)\times U(1)$ $biunitary$ transformation, i.e.
\begin{equation}
A= {\rm diag}(a,a,a_3)\;\;,\;\; B= U_2 V \tilde U_2^\dagger ~{\rm diag}(b,b,b_3)~ \tilde U_2 V^\dagger U_2^\dagger\;,
\end{equation}
where $U_2$ and $\tilde U_2$ have a $U(2)$ block in upper left corner and a phase in the (33) position.
Since any matrix can be diagonalized by a biunitary transformation, the upper left block of $V$ can be brought
to a diagonal form, i.e. $V_{12}'=V_{21}'=0$. The unitarity of $V'$ then requires $V_{13}'=V_{31}'=0$
and $\vert V_{11}' \vert =1$ (or $V_{23}'=V_{32}'=0$ and $\vert V_{22}' \vert =1$ ),
\begin{equation}
V'= \left( \matrix{ e^{i \theta }  & 0& 0 \cr
             0 & V_{22}' & V_{23}' \cr
             0 & V_{32}' & V_{33}'}
\right)\;.
\end{equation}
This results in $b_{12}=b_{13}=0$.
We therefore see that the residual symmetry allows us to eliminate two of the physical phases. 
The remaining CP phases are
\begin{equation}
\phi_0' \;,\;\phi_1 \;.
\end{equation}
Finally, if two eigenvalues of $C$ are also degenerate, $\phi_0'=0$  and the only physical phase is 
\begin{equation}
\phi_1 \;.
\end{equation}
Note that in this case Tr$[A,B]C \propto \sin \phi_1$. 

Now let us briefly discuss the case of three degenerate eigenvalues. If $A \propto $ {\bf I}, 
the residual symmetry is $U(3)$ which allows us to diagonalize another matrix, say $B$. 
Then the only invariant   phase is 
\begin{equation}
\phi_0' 
\end{equation}
and $J_{BC}\propto \sin\phi_0'$. It is clear that if either $B$ or $C$ have degenerate eigenvalues,
no CP violation occurs. 

These results can also be obtained by a naive parameter counting.
$U(2)$ has an extra phase parameter compared to $U(1)\times U(1) $,  so enlarging
the residual symmetry from $U(1)\times U(1) $ to $U(2)$ will reduce the number of physical phases by one.
Similarly, $U(3)$ will allow us to eliminate three more phases compared to $U(1)\times U(1) \times U(1)$.

The number of the physical phases also reduces if some of the mixings are zero.
The mixing angles are defined through the following parametrization 
of a unitary matrix $V$ (up to a phase transformation) \cite{Hagiwara:pw}:
\begin{equation}
V= \left( \matrix{ \mathcal C_{12} \mathcal C_{13} & \mathcal S_{12} \mathcal C_{13} & 
                   \mathcal S_{13} e^{-i \delta_{13}} \cr
  -\mathcal S_{12} \mathcal C_{23}- \mathcal C_{12} \mathcal S_{23} \mathcal S_{13} e^{i \delta_{13}} &  \mathcal C_{12} \mathcal C_{23} -\mathcal S_{12}\mathcal S_{23} \mathcal S_{13} e^{i \delta_{13}} & 
 \mathcal S_{23} \mathcal C_{13} \cr
 \mathcal S_{12} \mathcal S_{23}- \mathcal C_{12} \mathcal C_{23} \mathcal S_{13} e^{i \delta_{13}} &
 -\mathcal C_{12} \mathcal S_{23} -\mathcal S_{12}\mathcal C_{23} \mathcal S_{13} e^{i \delta_{13}} &
  \mathcal C_{23} \mathcal C_{13}  }
\right)\;,
\end{equation}
where  $\delta_{13}$ is a phase; $\mathcal S_{ij}=\sin \theta_{ij}$, $\mathcal C_{ij}=\cos \theta_{ij}$,
and $\theta_{12}, \theta_{13}, \theta_{23}$ are the mixing angles.
It is easy to see that 
one vanishing mixing angle annuls one phase, say the CKM--type phase $\phi_0$. Another vanishing mixing
eliminates $\phi_0'$ which is equivalent to saying $\phi_1 +\phi_2 +\phi_3=0$.
If both of these mixings are in the same matrix, say $V$, then this implies that
two of the elements $\{  b_{12},b_{13},b_{23}  \}$ vanish, so again two of the physical
phases disappear. 
Further, the third zero mixing would eliminate  another phase.
The next step, however, is nontrivial. If one matrix contains three zero mixings and the other -- one,
then no CP violation is possible. 
Indeed, this means that two matrices, e.g. $A$ and $B$, are diagonalizable
simultaneously so that  all $K$--invariants vanish, whereas the $J$--invariants vanish due to a single
zero mixing. On the other hand, if $V$ and $U$ have two zero mixings each, then CP violation is still 
possible. As mentioned above, in this case two of $\{  b_{12},b_{13},b_{23}  \}$ and 
two of $\{  c_{12},c_{13},c_{23}  \}$ vanish, so that one can have 
\begin{eqnarray}
{A= \left(      \matrix{a_1 & 0 & 0 \cr
                       0 & a_2 & 0 \cr
                       0 & 0  & a_3  }   \right) \;,\;
B=\left(      \matrix{b_{11} &  b_{12} &  0 \cr
                        b_{12}^* & b_{22} &  0 \cr
                        0 & 0  & b_{33}  }   \right) \;,\;
C=\left(      \matrix{c_{11} & c_{12} & 0   \cr
                      c_{12}^* & c_{22} & 0 \cr
                      0  &  0  & c_{33}  }   \right) \;.}  
\end{eqnarray}
The surviving invariant phase is $\phi_3={\rm Arg}(b_{12}c_{12}^*)$ and 
Tr$[A,B]C \propto \sin \phi_3$. If the non--vanishing off--diagonal entries are misaligned,
CP is conserved. No CP violation can occur if five of the mixing angles are zero.

\subsection{Generalization to more than three matrices.}

Suppose we have $N$ hermitian matrices $H_1, H_2,..H_N$ with the same transformation properties,
$H_i \rightarrow U_L H_i U_L^\dagger$. 
How can the results of the previous subsection be generalized for this case?

The generalization is quite straightforward.
Using the unitary freedom, we bring $H_1$ to the diagonal form: 
\begin{eqnarray}
H_1= \left(      \matrix{(H_1)_1 & 0 & 0 \cr
                       0 & (H_1)_2 & 0 \cr
                       0 & 0  & (H_1)_3  }   \right) \;,\;
H_2=\left(      \matrix{(H_2)_{11} &  (H_2)_{12} &  (H_2)_{13} \cr
                        (H_2)_{12}^* & (H_2)_{22} &  (H_2)_{23} \cr
                        (H_2)_{13}^* & (H_2)_{23}^*  & (H_2)_{33}  }   \right) \;,\; ...
\end{eqnarray}
The reparametrization invariant phases can be constructed by taking cyclic products of the 
elements of the same matrix or by taking products of elements  in the same positions
in different matrices.
 In the non--degenerate case, the $3N-5$ independent phases can be chosen as
\begin{equation}
\phi_0= {\rm Arg}\left[ (H_2)_{12} (H_2)_{13}^* (H_2)_{23}   \right]\;\;,\;\;
\phi_i^a= \epsilon_{ijk} {\rm Arg} \left[ (H_2)_{jk} (H_a)_{jk}^*  \right]\;,\; a=3...N\;.
\label{phases1}
\end{equation}
Any other physical phase can be expressed in terms of these basis phases. The corresponding weak basis invariants are
\begin{equation}
J_{H_1 H_2} \;\;,\;\; K_{H_1 H_2 H_a}(p,q,r)  \;,\;a=3...N\;,
\label{basis1}
\end{equation}
with $( p,q,r)=\{ (1,1,1); (2,1,1); (1,2,1) \}$.

In the degenerate case, the discussion of the previous subsection equally applies.
Any additional $U(2)$ symmetry, i.e. the presence of two degenerate eigenvalues, 
eliminates one physical phase  which can be taken to be the CKM--type phase for this matrix. 
An extra $U(3)$ eliminates three physical phases, for instance $\phi^a_{1,2,3}$. 
A vanishing mixing angle typically entails one vanishing phase, yet there are subtleties
discussed  above.

The necessary and sufficient conditions for CP conservation  can  be written as 
\begin{equation}
{\rm  Im Tr}[H_i,H_j]^3={\rm  Im Tr}H^p_{[i}H^q_j H^r_{k]}=0
\end{equation}
for any $i,j,k$ and $p,q,r$. 
Here the square brackets denote antisymmetrization with respect to the indices.
These conditions amount to 
\begin{equation}
{\rm Arg}\left[ (H_\alpha)_{12} (H_\alpha)_{13}^* (H_\alpha)_{23}   \right]=
 \epsilon_{ijk} {\rm Arg} \left[ (H_\alpha)_{jk} (H_\beta)_{jk}^*  \right]=0 \; {\rm mod}\; \pi
\label{phases2}
\end{equation}
for all $\alpha,\beta,$ and $i$.

One may wonder whether it is possible to construct CP--odd $K$-- type invariants with more than three matrices under the trace. This is certainly possible, yet they will be functions of the basic
reparametrization invariant phases (\ref{phases1}) or, in a more general case,
(\ref{phases2}), and thus will not provide independent CP violating quantities.

\section{The Minimal Supersymmetric Standard Model.}

The general technology of the previous section can be applied (with some reservations) to the case
of the Minimal Supersymmetric Standard Model (MSSM).
The MSSM has a number of flavor structures  which transform under the   
$U(3)_{\hat Q_L} \times U(3)_{\hat U_R} \times U(3)_{\hat D_R}$ symmetry.
In particular, the transformation properties are given by
\begin{eqnarray}
 Y^u &\longrightarrow& U_L  Y^u U_{u_{_R}}^\dagger \;,\nonumber\\
 Y^d &\longrightarrow& U_L  Y^d U_{d_{_R}}^\dagger \;,\nonumber\\
 A^u &\longrightarrow& U_L  A^u U_{u_{_R}}^\dagger \;,\nonumber\\
 A^d &\longrightarrow& U_L  A^d U_{d_{_R}}^\dagger \;,\nonumber\\
 M^{2 q_{_L}} &\longrightarrow& U_L M^{2 q_{_L}}  U_{L}^\dagger \;,\nonumber\\
 M^{2 u_{_R}} &\longrightarrow& U_{u_{_R}} M^{2 u_{_R}}  U_{u_{_R}}^\dagger \;,\nonumber\\
 M^{2 d_{_R}} &\longrightarrow& U_{d_{_R}} M^{2 d_{_R}}  U_{d_{_R}}^\dagger \;.
\end{eqnarray}
To construct the weak basis invariants, it is necessary  to identify hermitian objects which
transform under one of the unitary groups. These are given in Table 1.
Not all of them are, however, independent. In particular, only three matrices out of 
\begin{equation}
A^u A^{u\dagger} \;,\; A^{u\dagger} A^{u} \;,\; A^u Y^{u\dagger}+{\rm h.c.} \;,\;
A^{u\dagger} Y^{u}+{\rm h.c.}
\end{equation}
contain independent phases in the off-diagonal elements.
 This can be seen as follows. Let us go over to the basis where $Y^u$ is diagonal,
$Y^u \rightarrow U_L  Y^u U_{u_{_R}}^\dagger ={\rm diag}(m_u,m_c,m_t)/v_2$.
Given $A^u A^{u\dagger}$ and   $ A^{u\dagger} A^{u}$, we can find $A^u$ up to a phase transformation.
Indeed,  $A^u A^{u\dagger}$ and $ A^{u\dagger} A^{u}$ fix the diagonalization matrices of $A^u$:
\begin{eqnarray}
 A^u A^{u\dagger} &\longrightarrow& \tilde U_L  A^u  A^{u\dagger} \tilde U_{L}^\dagger \;\;\; =
   {\rm diag} (a_1^{u2},a_2^{u2},a_3^{u2})\;,\nonumber\\
  A^{u\dagger} A^u &\longrightarrow& \tilde U_{u_{_R}}    A^{u\dagger} A^u  \tilde U_{u_{_R}}^\dagger =
   {\rm diag} (a_1^{u2},a_2^{u2},a_3^{u2})\;, 
\end{eqnarray}
so that $A^u$ is given by 
\begin{equation}
A^u =\tilde U_{u_{_R}}^\dagger {\rm diag} (a_1^{u},a_2^{u},a_3^{u}) \tilde U_L \;.
\end{equation}
Note that both  $\tilde U_{u_{_R}}$ and $\tilde U_L$ are only defined up to a diagonal phase transformation,
\begin{equation}
\tilde U_{u_{_R}} \sim {\rm diag}(e^{i\delta_1}, e^{i\delta_2},e^{i\delta_3}) ~\tilde U_{u_{_R}}\;\;,\;\;
\tilde U_{L} \sim {\rm diag}(e^{i\phi_1}, e^{i\phi_2},e^{i\phi_3})~ \tilde U_{L} \;.
\end{equation}
This introduces a phase ambiguity in the matrix elements of $A^u$. 
A phase transformation with  $\delta_i  = \phi_i $ leaves  $A^u$  intact, 
whereas that with  $\delta_i  = -\phi_i $ changes it. This remaining ambiguity is eliminated by
fixing $A^u Y^{u\dagger}+{\rm h.c.}$ such that $A^u$ and $A^{u\dagger} Y^{u}+{\rm h.c.}$ can be 
determined unambiguously. This can also be understood by  parameter counting:
$A^u$ has nine phases which, in the non--degenerate case, can be found from the nine phases of the three
hermitian matrices $A^u A^{u\dagger}$, $ A^{u\dagger} A^{u}$, and $A^u Y^{u\dagger}+{\rm h.c.}$
Of course, similar considerations apply to $A^{d\dagger} Y^{d}+{\rm h.c.}$

This argument can be generalized to an arbitrary number of generations $N$.  $N^2$ phases of $A^u$
can be found via $N(N-1)$ phases of  
$A^u A^{u\dagger}$ and $ A^{u\dagger} A^{u}$, 
and $N$ phases of $A^u Y^{u\dagger}+{\rm h.c.}$
Although $A^u Y^{u\dagger}+{\rm h.c.}$ has $N(N-1)/2$ phases, 
only $N$ of them are independent and  correspond to the residual phase freedom 
with $\tilde U_L = \tilde U_{u_{_R}}^\dagger ={\rm diag }(e^{i\delta_1},..,e^{i\delta_N} )$.
This, however, does not work if $N>N(N-1)/2$ in which case $\delta_i$ cannot be determined unambiguously from 
the off--diagonal phases of $A^u Y^{u\dagger}+{\rm h.c.}$ So, for $N=1$ and $N=2$, additional information besides
the hermitian quantities is needed which can be, for instance, 
the anti--hermitian matrix $A^u Y^{u\dagger}-{\rm h.c.}$ 

Let us now identify the physical CP phases.
First of all, the physical phases must be invariant under the $U(1)_{\rm PQ}$ and 
$U(1)_{\rm R}$ symmetries. By an $R$--rotation, the gluino mass can be made real, so henceforth
the phases of the A--terms will be assumed to be relative to the gluino phase. Similarly, by a Peccei--Quinn
transformation the $B\mu$ term can be made real. Thus we have three CP phases in the flavor independent objects
$\mu$,$M_1$, and $M_2$. The phases of the flavor--dependent objects can be easily counted:
out of the original 45 phases of the flavor objects 17 can be eliminated
by the $U(3)^3$ symmetry (one of the $U(1)$ transformations leaves all flavor structures intact),
leaving 28 physical phases. 
These can be expressed in terms of the reparametrization invariant phases of the hermitian matrices.
The three columns of Table 1 form three separate sequences.
Using the unitary freedom, the first matrix in each column  can be made diagonal.
Then the strategy of section 2 can be applied. Each column has $3N-5$ physical phases, where 
$N$ is the number of matrices in the column. However, as I argued above, the two matrices   
$A^{u\dagger} Y^{u}+{\rm h.c.}$ and $A^{d\dagger} Y^{d}+{\rm h.c.}$ are not independent.
In particular, that means that, in the second column, the CKM--type phase for 
$A^{u\dagger} Y^{u}+{\rm h.c.}$ is not an independent phase, whereas the phase differences of
the off--diagonal elements of $A^{u\dagger} Y^{u}+{\rm h.c.}$ and $ M^{2 u_{_R}}   $,
and $A^{u\dagger} Y^{u}+{\rm h.c.}$ and $A^{u\dagger} A^u $ are independent.
Therefore, instead of 7 phases in the second column we  should  only count 6, and similarly for the third column.
Thus, we end up with 16+6+6=28 physical independent phases, as expected.

\begin{table}
\begin{center}
\begin{tabular}{|c|c|c|}
\hline
  $U(3)_{\hat Q_L}$   & $U(3)_{\hat U_R} $  & $ U(3)_{\hat D_R} $ \\
\hline
\hline
$Y^u Y^{u\dagger}   $ & $ Y^{u\dagger} Y^u $ &  $ Y^{d\dagger} Y^d $ \\
$Y^d Y^{d\dagger}   $ & $ M^{2 u_{_R}}   $   &  $ M^{2 d_{_R}}   $ \\
$M^{2 q_{_L}}$        &    $A^{u\dagger} A^u $ &  $A^{d\dagger} A^d $ \\
$A^u A^{u\dagger}$    & $ A^{u\dagger} Y^{u}+{\rm h.c.} $ & $ A^{d\dagger} Y^{d}+{\rm h.c.} $  \\
$A^d A^{d\dagger}$    & $  $ & $ $ \\
$A^u Y^{u\dagger}+{\rm h.c.}$ & $ $ & $ $ \\
$A^d Y^{d\dagger}+{\rm h.c.}$ & $ $ & $ $ \\
\hline
\end{tabular}
\end{center}
\caption{Hermitian objects of the MSSM  transforming under the unitary flavor symmetries. }
\label{table1}
\end{table}

In the basis where $Y^u Y^{u\dagger}$, $Y^{u\dagger} Y^u$, and $Y^{d\dagger} Y^d$ are diagonal, 
the 28 independent physical phases can be chosen as follows:
\begin{eqnarray}
&&\phi_0={\rm Arg}\left[ (Y^d Y^{d\dagger})_{12} (Y^d Y^{d\dagger})_{13}^* (Y^d Y^{d\dagger})_{23}   \right] \;,\nonumber\\
&& \phi_i^a= \epsilon_{ijk} {\rm Arg} \biggl[ (Y^d Y^{d\dagger})_{jk} (F^a)_{jk}^*  \biggr] \;,\; \nonumber\\
&& \chi_i^a= \epsilon_{ijk} {\rm Arg} \biggl[ ( A^{u\dagger} Y^u +{\rm h.c.})_{jk} (G^a)_{jk}^*  \biggr] \;,\; \nonumber\\      
&& \xi_i^a= \epsilon_{ijk} {\rm Arg} \biggl[ ( A^{d\dagger} Y^d +{\rm h.c.})_{jk} (H^a)_{jk}^*  \biggr] \;,\;      
\label{allphases}
\end{eqnarray}
where
\begin{eqnarray}
&&F^a=  \{ M^{2 q_{_L}},A^u A^{u\dagger},A^d A^{d\dagger},A^u Y^{u\dagger}+{\rm h.c.},A^d Y^{d\dagger}+{\rm h.c.}  \} \;, \nonumber\\
&&G^a=  \{ M^{2 u_{_R}}, A^{u\dagger} A^u  \} \;, \nonumber\\
&&H^a=  \{ M^{2 d_{_R}}, A^{d\dagger} A^d  \} \;.
\end{eqnarray}
Here I have assumed that there are no degenerate eigenvalues and the mixing angles are non--zero.
The corresponding weak basis CP--odd invariants are given by Eq.(\ref{basis1}).

In the degenerate case, i.e.  when some mixing angles are zero and/or there are degenerate eigenvalues, 
the situation becomes more complicated. In particular, the intrinsically non--hermitian objects such as 
$A^\alpha$ may have CP--phases which cannot be ``picked up'' by the hermitian quantities in the degenerate case.
This occurs, for instance, when only a 2$\times$2 block of $A^\alpha$ is non--zero such that we effectively deal
with two generations.
Another example is the case when $A^u$ and $Y^u$  can be diagonalized simultaneously. Then, in the basis where they are 
diagonal, the reparametrization invariant CP--phases are 
\begin{equation}
\rho_i^u= {\rm Arg} (A_{ii}^u Y_{ii}^{u *}) 
\label{rho}
\end{equation}
and similarly for $A^d$.
On the other hand, the hermitian matrices $A^u A^{u\dagger},A^{u\dagger} A^u $ and 
$A^u Y^{u\dagger}+{\rm h.c.}$ are diagonal (and real), so there are no CP phases
associated with them.  The CP violating invariants corresponding to the phases $\rho_i^u$ are based
on the anti--hermitian matrices:
\begin{equation}
L_{A^u Y^u}(p)={\rm Im Tr}\left[ (A^u Y^{u\dagger})^p- {\rm h.c.}  \right]\;,
\end{equation}
where $p$ is an integer. For $p=1$, this becomes 
\begin{equation}
L_{A^u Y^u}(1)=2 \sum_i \vert A_{ii}^u m_i^u/v_2    \vert \sin\rho_i^u \;.
\end{equation}
Quantities of the type ${\rm Im Tr} (A^{\alpha\dagger} Y^{\alpha})^p$ do not provide independent
CP--violating invariants.
Further, if all $A_{ij}^u$ are non--zero in the basis where $Y^u$ is diagonal, $\rho_i^u$ are not independent and 
are functions of the phases (\ref{allphases}). 

The necessary and sufficient conditions for CP conservation are
\begin{eqnarray}
&& {\rm  Im Tr}[M_i,M_j]^3={\rm  Im Tr}M^p_{[i}M^q_j M^r_{k]}=0 \;, \nonumber \\
&& {\rm Im Tr}  (A^\alpha Y^{\alpha\dagger})^p =0
\end{eqnarray}
for any $i,j,k$ and $p,q,r$; $\alpha=\{ u,d \}$. Here $M_i$ are hermitian matrices belonging to the same column of Table 1
and these conditions are to be satisfied for each column. 
In addition, one, of course, has to require that the
gaugino and the $\mu$-term phases vanish. In this case, the full MSSM will conserve CP. 
 If all of the physical phases (\ref{allphases}) and (\ref{rho}) are much smaller than one,  
CP is an approximate symmetry
(yet, this is unrealistic
as the CKM phase is of order one  experimentally). In terms of basis--independent quantities,
this implies that, when the mixing
angles and the eigenvalues are fixed, the $J$-, $K$-, and $L$-invariants
are close to zero (in the appropriate units).

If we are to require that  the CKM phase be the only source of CP violation, all invariants apart from 
${\rm  Im Tr}[Y^u Y^{u\dagger},Y^d Y^{d\dagger}]^3$ must vanish.
This occurs, for instance, when all SUSY flavor structures are proportional to the unit matrix and the A--terms are real.
It is important to note that it is not sufficient to require that the SUSY flavor structures be
real in some basis because the presence of CP phases in the Yukawas can make some of the invariants,
apart from the Jarlskog one, non--zero. For example, when $Y^u Y^{u\dagger}$ is diagonal,
$Y^d Y^{d\dagger}=V{\rm diag}(m_d^2,m_s^2,m_b^2)V^\dagger/v_1^2$, and $M^{2 q_{_L}}$ is real,
the invariant phases $\phi_i^1$ and ${\rm  Im Tr}[Y^u Y^{u\dagger},Y^d Y^{d\dagger}] M^{2 q_{_L}}$ are  nonzero. 
This creates certain difficulties for realistic string models with low energy supersymmetry \cite{Abel:2001cv}.
The reason is that such models generally predict non--universal A--terms. Then even if the SUSY breaking
F--terms are real, the complex phases of the type (\ref{rho}) are generated by the basis transformation
which brings the Yukawa couplings to the diagonal form. Equivalently, this means that some of the $L$--
or $K$--invariants are non--zero.  As a result, large electric dipole moments of fermions
are induced, in conflict with experiment.

Finally, it should be noted that after the electroweak symmetry breaking a number of corrections to
the Lagrangian (\ref{l}) will appear.  In particular, $M^{2 q_{_L}}$ will split into 
$M^{2 u_{_L}}$ and $M^{2 d_{_L}}$ due to the isospin breaking corrections.
Since no additional sources of CP violation arise in this process,
the consequent CP phases will be functions of the original phases 
(\ref{allphases}) and (\ref{rho}).

\subsection{Examples.}

Let us consider a few phenomenological examples.

{\bf i. Kaon mixing.}  A first example is a supersymmetric contribution to the $K -\bar K$ mixing.
This is conveniently expressed in terms of the mass insertions \cite{Hall:1985dx}. Suppose that we work
in the super--CKM basis, i.e. the basis where the quark mass matrices are diagonal,
and that the only non--vanishing mass insertion is $(\delta_{LL}^d)_{12}\equiv ( M^{2 d_{_L}})_{12}/\tilde m^2$
where $\tilde m^2$ is the average squark mass. 
Then 
the gluino--mediated contribution to the $\varepsilon_K$ parameter  
is usually written as \cite{Gabbiani:1996hi}
$ (\varepsilon_K)_{_{\rm SUSY}} \propto  {\rm Im} (\delta_{LL}^d)_{12}^2\;.$ 
Clearly, this result is $not$ rephasing invariant.
The super--CKM basis is defined only up to a phase transformation, which physics should be independent of. 
It can be shown that, in fact\footnote{This follows from the penguin dominance in the kaon decay 
amplitude.},
\begin{equation}
(\varepsilon_K)_{_{\rm SUSY}} \propto  {\rm Im} \Bigl[ (\delta_{LL}^d)_{12} V_{21} V_{22}^*  
\Bigr]^2\;.
\end{equation}
As a result, the $K -\bar K$ mixing constrains the invariant 
quantity $(\delta_{LL}^d)_{12} V_{21} V_{22}^*$ and even a $real$ mass insertion can lead 
to CP violation.
Ref.\cite{Gabbiani:1996hi} assumes a special form of the CKM matrix in which $V_{21} V_{22}^*$
is real.  Note, however, that, unlike in the Standard Model, different CKM ``conventions''
are $not$ $physically$ $equivalent$. This is due to the presence of the additional invariant phases
(\ref{phases1}). In other words, diagonalization of the Yukawa matrices leads to the CKM matrix
in some generic phase convention. To bring it to a special form, one rotates the quark and 
squark fields simultaneously thereby inducing CP--phases in $M^{2 d_{_L}}$
even if it was real initially. Equivalently, the $K$-invariants such as
${\rm Im Tr}[Y^d Y^{d\dagger},Y^u Y^{u\dagger}]M^{2 d_{_L}}$ may not vanish even if
$M^{2 d_{_L}}$ is real.
Thus, in all phenomenological analyses the definition
of the super-CKM basis must be supplemented with the specification of the phase convention  of
the CKM matrix.

The reparametrization invariant phases can be identified as follows.
By choosing an appropriate $U_L$, let us go to the basis where
\begin{equation}
Y^d Y^{d\dagger}={\rm diag}(m_d^2,m_s^2,m_b^2)/v_1^2 \;\;,\;\;
Y^u Y^{u\dagger}=V^\dagger {\rm diag}(m_u^2,m_c^2,m_t^2)V /v_2^2 \;,
\end{equation}
where $V$ is the CKM matrix.
In this basis  $M^{2 d_{_L}}$ is the same as in the super--CKM basis.
The reparametrization invariant CP phases are the relative phases between
the matrix elements of  $Y^u Y^{u\dagger}$ and $M^{2 d_{_L}} $ in the same position. In particular,
the relevant to the $K-\bar K$ mixing invariant  phase is
\begin{equation} 
\delta = {\rm Arg}\left(  \sum_i \vert m_i^{u}\vert^2 V_{i1} V_{i2}^*  \right) \left( M^{2 d_{_L}} \right)_{12} \;.
\end{equation}
Here I have used the absolute values of the masses to stress their invariance with respect to the phase transformations.
If $(\delta_{LL}^d)_{12}$ is the only non--zero mass insertion, there is only one additional
invariant phase -- the CKM phase. Then, $\varepsilon_K$ is a function of these two phases. 
Specifically, the relevant CP phase is expressed as
\begin{equation}
{\rm Arg} \Bigl[   (\delta_{LL}^d)_{12} V_{21} V_{22}^*    \Bigr]=
\delta - \phi (\delta_{_{\rm CKM}})\;,
\end{equation}
where $\phi (\delta_{_{\rm CKM}})$ is an invariant phase defined by
\begin{equation}
\phi (\delta_{_{\rm CKM}})={\rm Arg} \left[ 1+ {\vert m_u \vert^2 V_{11} V_{12}^* \over
\vert m_c \vert^2 V_{21} V_{22}^* }   + {\vert m_t \vert^2 V_{31} V_{32}^* \over
\vert m_c \vert^2 V_{21} V_{22}^* }
   \right]\;.
\end{equation}

The physical phases can be written in terms of the basis--independent quantities.
In particular, if $(\delta_{LL}^d)_{12}$ is the only non--zero mass insertion  then 
\begin{equation}
{\rm Im Tr}[Y^d Y^{d\dagger},Y^u Y^{u\dagger}]M^{2 d_{_L}}=
2 {m_s^2-m_d^2 \over v_1^2 v_2^2} \biggl\vert  \sum_i \vert m_i^{u}\vert^2 V_{i1} V_{i2}^*   
  \left( M^{2 d_{_L}} \right)_{12}  \biggr\vert ~\sin\delta \;.
\end{equation}
In a more general case, the invariant phases and the magnitudes of the matrix elements of $M^{2 d_{_L}}$
can be 
found via 3 CP--violating and 6 CP--conserving weak basis invariants
\begin{eqnarray}
&& {\rm Im Tr}[Y^d Y^{d\dagger},Y^u Y^{u\dagger}]M^{2 d_{_L}} \;,
   {\rm Im Tr}[(Y^d Y^{d\dagger})^2,Y^u Y^{u\dagger}]M^{2 d_{_L}} \;,
   {\rm Im Tr}[Y^d Y^{d\dagger},(Y^u Y^{u\dagger})^2]M^{2 d_{_L}} ,\nonumber\\
&& {\rm  Tr~} Y^u Y^{u\dagger}M^{2 d_{_L}} \;,\; {\rm  Tr~} (Y^u Y^{u\dagger})^2M^{2 d_{_L}} \;,\;
   {\rm  Tr~} Y^u Y^{u\dagger}(M^{2 d_{_L}})^2 \;,\; \nonumber\\
&& {\rm  Tr~} Y^d Y^{d\dagger}M^{2 d_{_L}} \;,\; {\rm  Tr~} (Y^d Y^{d\dagger})^2M^{2 d_{_L}} \;,\;
   {\rm  Tr~} Y^d Y^{d\dagger}(M^{2 d_{_L}})^2 \;. 
\end{eqnarray}

{\bf ii. EDMs.} Another example is a SUSY contribution to the electric dipole moments of the quarks.
For the down quark, the relevant gluino--mediated contribution is typically written as \cite{Gabbiani:1996hi}
\begin{equation}
(d_{\rm d})_{_{\rm SUSY}} \propto   {\rm Im} (\delta_{LR}^d)_{11} M_3^* \;, 
\label{edm}
\end{equation}
with $(\delta_{LR}^d)_{11} \sim v_1 A^d_{11}/ \tilde m^2$ (omitting the $\mu$-term contribution).
Again, it is clear that this result is not rephasing invariant.

To rectify this problem, one may use the strategy advocated above.
For the purpose of illustation, let us assume the following simple form of $A^d$
in the super--CKM basis:
\begin{equation}
A^d= \left(      \matrix{ A^d_{11}& A^d_{12} & 0 \cr
                       0 & 0 & 0 \cr
                       0 & 0  & 0  }   \right) \;.
\end{equation}
The relevant physical phase  can be expressed via the hermitian quantities of Table 1, column 3:
\begin{equation}
\xi= {\rm Arg}\left( A^{d\dagger}Y^d + {\rm h.c.}\right)_{12} \left( A^{d \dagger} A^d\right)_{12}^*=
 {\rm Arg~} A^{d}_{11}m_d^* \;.
\end{equation}
Since $m_d$ has the same transformation properties under the left and right rephasings as $A^d_{11}$,
this expression is manifestly reparametrization invariant.
The corresponding weak basis invariant is 
${\rm Im Tr}[Y^{d\dagger} Y^d, ( A^{d\dagger}Y^d + {\rm h.c.})]A^{d \dagger} A^d   \propto \sin\xi$.
Thus, Eq.(\ref{edm}) is to be modified as
\begin{equation}
(\delta_{LR}^d)_{11} \longrightarrow  \Bigl\vert (\delta_{LR}^d)_{11} \Bigr\vert e^{i \xi} \;.
\end{equation}
Clearly, if $m_d=0$, the SUSY EDM contribution 
vanishes.

The invariant phase associated with $A^d_{12}$ can similarly be found from 
$A^{d}Y^{d\dagger} + {\rm h.c.}$ and $Y^{u } Y^{u\dagger}$.
It is important to note that if $A^d_{12}$ vanishes, $A^d$ and $Y^d$ are diagonal simultaneously,
and the phase of $A^d_{11}$ $cannot$ be extracted from hermitian quantities. In this case, 
the relevant phase is given by Eq.(\ref{rho}) and the corresponding weak basis invariant is
$L_{A^d Y^d}(1)$.

\section{Summary.}

In this work, I have studied the CP--odd  weak basis invariants in supersymmetric models and the 
corresponding reparametrization invariant CP phases. 
I have shown that, 
in SUSY models, a new class of CP--odd invariants, not expressible in terms of the Jarlskog--type
invariants, appears. 
I have also obtained basis independent conditions for CP conservation and clarified the issues of
rephasing invariance of observable quantities.
This work was supported by PPARC.


\begin{thebibliography}{99}

\bibitem{Jarlskog:1985ht}
C.~Jarlskog,
Phys.\ Rev.\ Lett.\  {\bf 55}, 1039 (1985);
Z.\ Phys.\ C {\bf 29}, 491 (1985). See also
I.~Dunietz, O.~W.~Greenberg and D.~d.~Wu,
Phys.\ Rev.\ Lett.\  {\bf 55}, 2935 (1985);
D.~d.~Wu,
Phys.\ Rev.\ D {\bf 33}, 860 (1986);
F.~J.~Botella and L.~L.~Chau,
Phys.\ Lett.\ B {\bf 168}, 97 (1986).
For more than three generations, see
J.~Bernabeu, G.~C.~Branco and M.~Gronau,
Phys.\ Lett.\ B {\bf 169}, 243 (1986);
M.~Gronau, A.~Kfir and R.~Loewy,
Phys.\ Rev.\ Lett.\  {\bf 56}, 1538 (1986).
For beyond the Standard Model, see e.g.
G.~C.~Branco and V.~A.~Kostelecky,
Phys.\ Rev.\ D {\bf 39}, 2075 (1989);
A.~I.~Sanda,
Phys.\ Rev.\ D {\bf 32}, 2992 (1985);
M.~Gluck,
Phys.\ Rev.\ D {\bf 33}, 3470 (1986);
A.~Mendez and A.~Pomarol,
Phys.\ Lett.\ B {\bf 272}, 313 (1991);
L.~Lavoura and J.~P.~Silva,
Phys.\ Rev.\ D {\bf 50}, 4619 (1994);
F.~J.~Botella and J.~P.~Silva,
Phys.\ Rev.\ D {\bf 51}, 3870 (1995);
J.~A.~Aguilar-Saavedra,
J.\ Phys.\ G {\bf 24}, L31 (1998).



\bibitem{Haber:1984rc}
For a review, see H.~E.~Haber and G.~L.~Kane,
Phys.\ Rept.\  {\bf 117}, 75 (1985).

\bibitem{Abel:2001vy}
S.~Dimopoulos and S.~Thomas,
Nucl.\ Phys.\ B {\bf 465}, 23 (1996).
See also 
D.~A.~Demir,
Phys.\ Rev.\ D {\bf 62}, 075003 (2000);
S.~Abel, S.~Khalil and O.~Lebedev,
Nucl.\ Phys.\ B {\bf 606}, 151 (2001).


\bibitem{Hagiwara:pw}
K.~Hagiwara {\it et al.}  [Particle Data Group Collaboration],
Phys.\ Rev.\ D {\bf 66}, 010001 (2002).

\bibitem{Abel:2001cv}
S.~Abel, S.~Khalil and O.~Lebedev,
Phys. Rev. Lett. {\bf 89}, 121601 (2002).

\bibitem{Hall:1985dx}
L.~J.~Hall, V.~A.~Kostelecky and S.~Raby,
Nucl.\ Phys.\ B {\bf 267}, 415 (1986).

\bibitem{Gabbiani:1996hi}
F.~Gabbiani, E.~Gabrielli, A.~Masiero and L.~Silvestrini,
Nucl.\ Phys.\ B {\bf 477}, 321 (1996).

\end{thebibliography}
\end{document}